\documentclass[a4paper,onecolumn,12pt,assc]{IEEEtranASSC}
\ifCLASSINFOpdf
   \usepackage[pdftex]{graphicx}
   \DeclareGraphicsExtensions{.pdf,.jpeg,.png}
\else
   \usepackage[dvips]{graphicx}
  \DeclareGraphicsExtensions{.eps}
\fi
\begin{document}
%
\title{Observations of the D/H ratio in Methane in the atmosphere of Saturn's moon, Titan - where did the Saturnian system form?}


\author{\IEEEauthorblockN{Lucyna Kedziora-Chudczer \IEEEauthorrefmark{1},
Jeremy Bailey \IEEEauthorrefmark{1} and
Jonathan Horner\IEEEauthorrefmark{1}}

\IEEEauthorblockA{\IEEEauthorrefmark{1}
Department of Astrophysics, School of Physics,
The University of New South Wales, Sydney, Australia 2052}
}

\maketitle

\begin{abstract} 
The details of the Solar system's formation are still heavily
debated. Questions remain about the formation locations of the giant planets, and the degree to
which volatile material was mixed throughout the proto-planetary system. One diagnostic which
offers great promise in helping to unravel the history of planet formation is the study of the
level of deuteration in various Solar system bodies. For example, the D/H ratio of methane in
the atmosphere of Titan can be used as a diagnostic of the initial conditions of the solar nebula
within the region of giant planet formation, and can help us to determine where Titan
(and, by extension, the Saturnian system) accreted its volatile material. The level of Titanian
deuteration also has implications for both the sources and long term evolution of Titan's
atmospheric composition. We present the results of observations taken in the 1.58 $\mu$m window
using the NIFS spectrometer on the Gemini telescope, and model our data using the VSTAR
line--by--line transfer model, which yields a D/H ratio for Titan's atmosphere of
($143\pm16)\times10^{-6}$ \cite{Bailey2012}. We are currently in the process of  modeling the 
Gemini high resolution GNIRS spectra using new sets of line parameters derived for methane in
the region between 1.2-1.7 $\mu$m \cite{Campargue2012}. 
\end{abstract} 



\section*{Introduction}

Over the years, our understanding of the formation of our Solar system has evolved
dramatically. For a long time, it was believed that our planetary system was formed by a chance
encounter between the Sun and another star, which passed sufficiently close to the Sun to draw
a lengthy tongue of material from it which then coalesced to form the planets (e.g.
\cite{Lyt36, Wolf64}). In recent decades, it has become widely accepted that our planetary
system instead formed from a dynamically cold disk\footnote{Objects moving on dynamically cold
orbits have very low orbital eccentricity and inclination. Therefore a dynamically cold disk is rather flattened and consists of the protoplanetary material in orbits of very low eccentricity. Objects in dynamically cold orbits typically only experience collisions at very low relative velocities. Such collisions tend to be constructive rather than destructive.} of dust and gas around the proto-Sun. By the early
1990s, it was thought that the process of planetary formation was essentially a leisurely,
gentle process that occurred at a fixed location within the proto-planetary disk (e.g.
\cite{Lissauer93}). 

It was soon realised that the planet formation process must be somewhat more complicated,
involving proto-planets and planets migrating over many astronomical units from their formation
location to their current locations. This sea-change in our understanding of planet formation
came about, in part, as a result of the discovery of the first planets found around Sun-like stars
(e.g. \cite{Mayor95}), which were found moving on totally unexpected orbits (Jupiter-size
planets orbiting far closer to their host stars than Mercury to our Sun). Within our own Solar
system, planet formation models featuring migration were found to be necessary in order to
explain various aspects of the distribution of small bodies in the outer Solar system including
the orbits of Pluto and the Plutinos (e.g. \cite{Malhotra93, Malhotra95}), the proposed Late
Heavy Bombardment of the inner Solar system (e.g. \cite{Gomes05}) and the populations of Jovian
and Neptunian Trojans (e.g. \cite{Morbi05, Lyk09}).

Although it is now widely accepted that the giant planets migrated significant distances before
reaching their current locations, the nature, range, and pace of that migration are still
heavily debated. Some models suggest a relatively sedate migration for the giant planets (e.g.
\cite{Lyk10}), whilst others suggest highly chaotic migration featuring significant mutual
scattering between the giant planets (e.g. \cite{Thom02}). Whilst it is likely that dynamical
studies of the small body populations in the outer Solar system may help to resolve the
unanswered question of the nature of planetary migration (e.g. \cite{LykHor10, JHASSC}), the
observation of isotopic abundances in the ices and atmospheres of objects in the outer Solar
system will provide an important independent test of their formation locations (e.g.
\cite{HMH07, HM08}). In particular, the study of the deuterium-to-hydrogen (D/H) ratio in the icy and
gaseous material in the outer Solar system may well provide a signpost to the formation
location of the bodies in question.

In this work, we present the first results of an observational program targeted at determining
the D/H ratio within methane in the atmosphere of Saturn's largest satellite, Titan. In the next section, we describe in more detail the current state of play in our understanding of the way in
which the D/H ratio varied through the proto-planetary disk, before briefly discussing the
origin of methane in Titan's atmosphere in the following section. Next we detail our observations
of Titan and describe our modelling of the Titanian spectrum. Finally, we
present our conclusions and discuss future work.

\section*{Deuterium and the formation of the Solar system}

Protium ($^{1}$H) and deuterium ($^{2}$H or D) are the only two stable isotopes of hydrogen.
The largest mass difference between two isotopes of any element leads to rather distinct
properties of hydrogen isotopologues (for example HDO and H$_{2}$O). Differences in
boiling point, vapour pressure and reactivity of compounds with different isotopes will result
in their having unique natural abundances. Once formed, a process of isotopic fractionation
(resulting from the different physical properties of the two isotopes) can lead to the physical
separation of isotopologues through evaporation, condensation, melting or diffusion. The D/H
ratio has therefore long been considered to be a useful diagnostic of the geological history of
rock formations, bodies of fluid and organic materials.

Deuterium formed during the Big Bang is subsequently destroyed in stars, and therefore its
abundance in the interstellar medium decreases with time. Currently, the D/H ratio in the local
interstellar medium is of the order of $15\times10^{-6}$ \cite{McCullough1992}, while the
initial value, immediately after the big bang, may well have been as high as $30\times10^{-6}$
\cite{Noter2012}. The D/H ratio in the protostellar solar nebula has been estimated using
measurements of D/H in the atmospheres of Jupiter and Saturn, because HD does not fractionate
during the transition from the molecular to metallic form. Observations from the Galileo probe
mass spectrometer provide a D/H value of $26\times10^{-6}$ \cite{Mahaffy1998}. 

Elsewhere in the Solar system, however, the situation is far from this clear cut. The D/H
values measured for a variety of Solar system objects vary significantly. The D/H of Vienna
Standard Mean Ocean Water (V-SMOW) is $156 \times 10^{-6}$, whilst comets 1P/Halley \cite{Eberhardt95},
C/1996 B2 Hyakutake \cite{BM98}, C/1995 O1 Hale-Bopp \cite{Meier98} and 8P/Tuttle \cite{VN08}
are enriched in deuterium by a factor of between 1.5 and three times that value. 

\begin{figure}[!hb]
\centering
\includegraphics[width=5in]{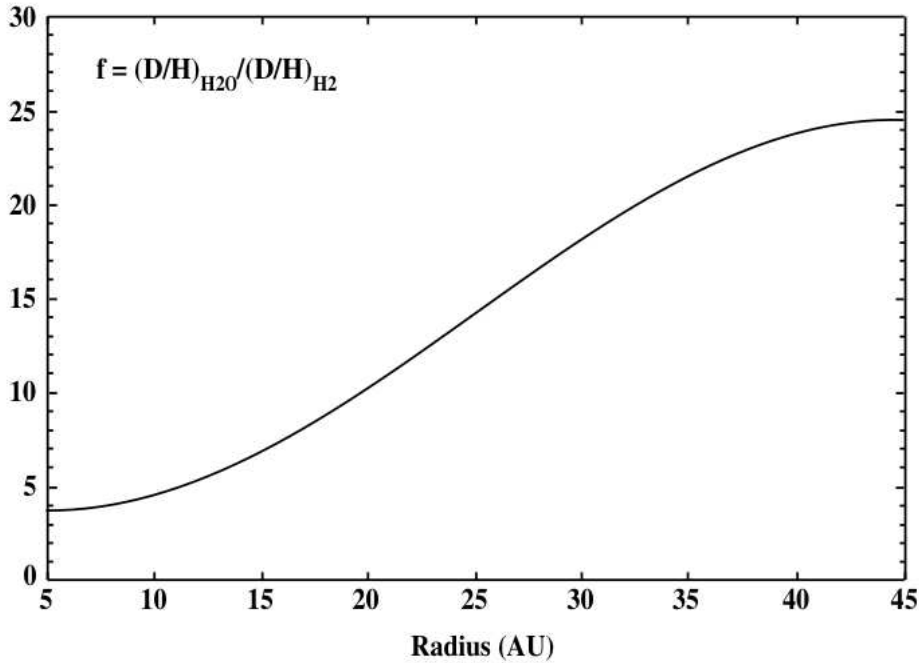}
\caption{The variation in the ratio of the deuteration in water to that in deuterated hydrogen, as a function of heliocentric distance, in the proto-planetary disk, taken from \cite{HMH07}. This illustrates how the deuteration of the volatiles incorporated into objects that formed beyond the ice-line will have been a strong function of their formation location.}
\label{DHWater}
\end{figure}

The enrichment of deuterium in water over that observed in H$_{2}$ is a direct result of equilibrium chemistry in the outer regions of the proto-planetary disk. In that disk, the isotopic fractionation of deuterium in water and deuterated hydrogen was governed by the reversible reaction shown in  the following equation \cite{HMH07}:
\begin{equation}
H_{2}O+HD\Longleftrightarrow HDO+H_{2}
\end{equation}
 At low temperatures, the sequestration of deuterium in water is favoured over its emplacement in deuterated hydrogen, and so there is a gradual enrichment of deuterium in water as a function of time. At greater heliocentric distances, it has been shown that the relative enrichment of deuterium in water is significantly greater than at small heliocentric distances, resulting in the model distribution of the variation in the deuteration of water compared to that in deuterated hydrogen shown in Figure \ref{DHWater}.

By contrast, the deuteration levels measured in meteorites are far closer to that measured in
the Earth's oceans (see e.g. Fig. 1 of \cite{Dr99}). That similarity has been used to suggest
that the bulk of the Earth's water was delivered by asteroidal, rather than cometary
bodies\footnote{The origin of the Earth's water has long been debated, with most modern
theories favouring an exogeneous source of our planet's oceans. We direct the interested reader
to section 4, "Planetary Features" of \cite{HJ10}, and references therein, for a detailed
discussion of the origin of Earth's water.} - with \cite{Robert2000} suggesting that, if the
carbonaceous chondrites were the main source of oceanic water, another suggested source
(the comets) could have contributed only a small fraction (less than 10\%). Following this
logic, it has often been assumed that the initial volatile budget of all the terrestrial
planets would have been identical to that of the Earth, although \cite{HMPJ09} showed that the
terrestrial planets would have experienced significantly different contributions of volatiles
from cometary and asteroidal bodies. 

A number of studies (e.g. \cite{Dr99, Mousis2000, HMH07}) have suggested that the D/H ratio
within ices accreted in the outer Solar system should vary significantly as a function of
heliocentric distance within the Solar nebula. The lowest D/H ratios would be found in objects
which accreted near the ice-line, the innermost location where water could condense as a solid,
and the largest D/H ratios would be found in volatiles that condensed beyond the current orbit
of Neptune. Figure~\ref{DHWater} shows the predicted enrichment in D/H for water over that in
deuterated hydrogen as a function of heliocentric distance in the outer Solar system. It is
clear that objects that formed at different heliocentric distances would have incorporated
volatiles containing vastly different amounts of deuterium. The less stirred the disk, the more
pronounced the variation in D/H as a function of heliocentric distance would be between one
body and the next (since increased stirring would doubtless bring icy bodies with high
deuteration to the inner reaches of the nebula). It is highly likely that the bulk of volatile
material accreted by a given object would be sourced from its immediate surroundings, however,
and so measurements of the deuteration of objects in the outer Solar system give us a tool by
which their formation location can be constrained.

\section*{Origin of CH$_4$ in the atmosphere of Titan}

In 1981, Prinn and Fegley \cite{Prinn1981} assumed that, in the primordial solar nebula, carbon
and nitrogen were entirely sequestered in CO and N$_{2}$. During the early stages of Saturn's
formation, a dense and warm sub-nebula developed, which enabled the conversion of CO to CH$_{4}$
and N$_{2}$ into NH$_{3}$. This methane and ammonia was accreted to form icy planetisimals, the
building blocks of Titan. By contrast, Mousis et al. \cite{Mousis2002} argued that such
conversion of CO and N$_{2}$ into methane and ammonia is unlikely to occur in conditions that
would have been prevalent in the Saturn sub-nebula, and that therefore CH$_{4}$ would have to be
present in the primary solar nebula in order to explains its accretion onto Titan.

According to the first model, methane becomes enriched in deuterium gradually when CH$_{3}$D in
vapour phase exchanges deuterium with hydrogen. After the temperature drops below 200K, this
process stops, but isotopic fractionation can continue through solar photolysis and outgassing
processes. 

The model put forward by Mousis et al. instead assumed that vaporisation of CH$_{4}$ happened
during the infall of icy planetisimals onto the protoplanetary disk, which allowed the exchange
of deuterium with hydrogen from the solar nebula. Methane was later released from the
subsurface layers to the atmosphere of Titan. The D/H ratio predicted from this model is in
good agreement with measurements without requiring  additional fractionation to have occurred.
Since methane undergoes solar photolysis by breaking into hydrocarbons, cryovolcanism was
proposed as its replenishment mechanism.

More recently Atreya et al. \cite{Atreya2006} have suggested that so-called 
``serpentinization'' reactions (i.e. hydrothermal reactions between silicates and water) in the interior of
Titan could generate CH$_4$. However, Mousis et al. \cite{Mousis2009} point out that this is
unlikely because it would imply a relatively low value for the D/H of the water involved
(similar to the V-SMOW value). They argue that water ice delivered to Titan should have
a higher D/H, similar to that measured in comets, and in the Enceladus plume \cite{Waite2009} and
that a primordial origin for Titan's methane is more likely.

\section*{D/H ratio measurements in Titan's atmosphere}

Two methods were used to determine the D/H ratio in Titan's atmosphere. In-situ measurements
were taken by the Cassini-Huygens probe \cite{Niemann2010}, and used the abundance of molecular
and deuterated hydrogen measured by the gas chromatograph mass spectrometer carried on-board,
yielding a D/H ratio of ($135\pm30 )\times10^{-6}$ for the satellite's atmosphere. 

Measurements of the D/H ratio can also be obtained using remote sensing
observations of the spectrum of Titan from the Cassini spacecraft and ground-based telescopes.
These observations are analysed using radiative transfer models, such
as the VSTAR model described in \cite{Bailey2012a}. Along with molecular nitrogen, which is the
most abundant gas in TitanÕs atmosphere, methane is a major component, comprising a few percent
of the atmosphere, as measured by mole fraction. The spectrum of methane has four vibrational
modes forming overlapping absorption bands in the infrared region, which become increasingly
complex towards higher energies. Measurements of Titan's D/H ratio from observations in the thermal infrared with the Cassini
CIRS instrument give a value of ($159\pm33)\times10^{-6}$ \cite{Nixon2012}

New, improved laboratory measurements of methane lines in the near infrared at temperatures relevant to cold regions of the
Solar system are now becoming available for use in radiative transfer models
\cite{Campargue2012}. These line lists can be used to model spectra of Titan observed with near
infrared spectrometers on ground-based telescopes.
The model shown
in Figure~\ref{figure1} yields a D/H ratio of ($143\pm16)\times10^{-6}$ \cite{Bailey2012} for
the atmosphere of Titan. De Bergh et al. \cite{DeBergh2012} used a different set of data from
KPNO/FTS with the improved list of methane lines from \cite{Wang2011}, and obtained a rather different result ($113\pm25)\times10^{-6}$. It is becoming clear that a number
of different datasets have to be analysed and modelled to resolve the differences between
estimates obtained from different methods.

\section*{Modelling of the 1.58 $\mu$m window of Titan's atmosphere}

\begin{figure}[!hb]
\centering
\includegraphics[width=6in]{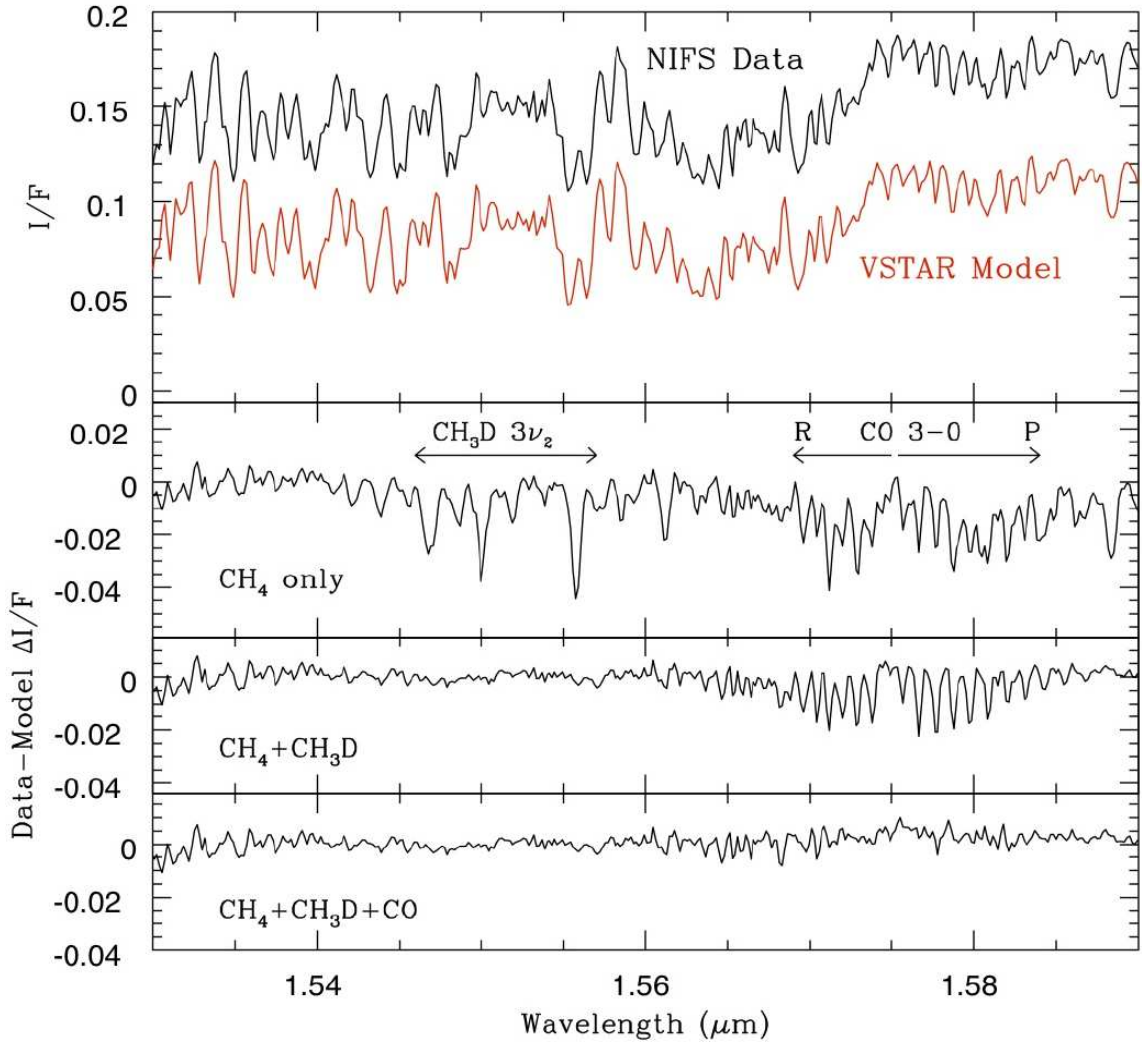}
\caption{NIFS spectrum in H-band region of the CH$_{3}$D absorptions compared with the VSTAR model (top panel). The residuals in panels below are described in more detail in the text and in \cite{Bailey2012}.}
\label{figure1}
\end{figure}

We obtained spectra of Titan in the J, H and K bands (1.1 to 
2.4 $\mu$m) with the Near-Infrared Integral Field Spectrometer (NIFS) on the
Gemini North 8m telescope. The NIFS spectrum
in the 1.53 to 1.59 $\mu$m region and its VSTAR model are reproduced from \cite{Bailey2012} in
Figure \ref{figure1}, where data and model are shown in the top panel. In the second panel from
the top, the residuals are plotted between the data and the model, which includes only CH$_{4}$
features. In the panel below, the absorption from the CH$_{3}$D are added, while in the bottom
panel the addition of CO 3-0 lines based on a mixing ratio of 50 ppmv completes the removal
of unaccounted residuals from this CO band. The details of the modelling techniques are
described in \cite{Bailey2011} and the parameters of the final model are presented
in Table 2 of \cite{Bailey2012}.  

As discussed in the previous section, the accuracy of the D/H measurements obtained from the near IR ground-based observations with NIFS is significantly higher than that of the in-situ Huygens probe measurements, or of the Cassini spacecraft results. An important advantage of ground-based observing is the ability to obtain spectra at higher resolution than that of space instruments. We are therefore
pursuing a program of observation of Titan and the methane-rich giant planets in the Solar
system with even higher resolution than the R $\sim$ 5000 of the NIFS observations. The aim is to use
the high-resolution spectra in conjunction with the latest methane spectral line data to
measure D/H for all these objects as accurately as is possible.

\begin{figure}[!hb]
\centering
\includegraphics[width=5in]{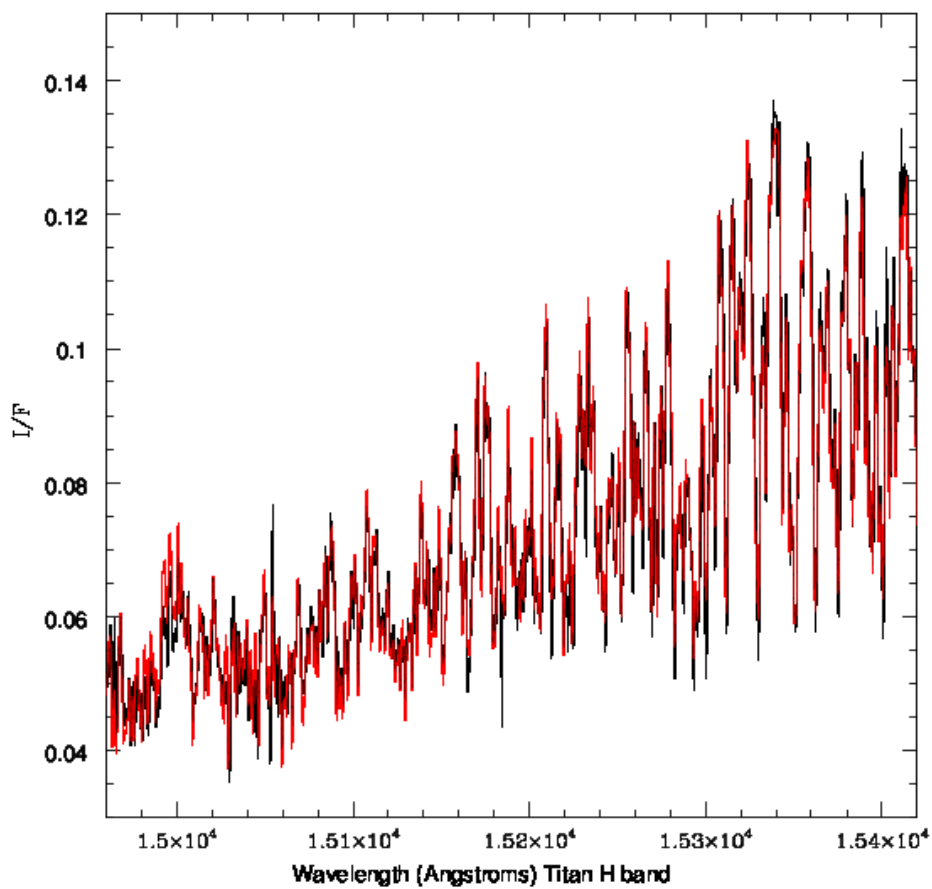}
\caption{GNIRS spectrum of Titan reflectance (I/F) plotted in black with the VSTAR model overlaid in red.}
\label{figure2}
\end{figure}

\begin{figure}[!hb]
\centering
\includegraphics[width=5in]{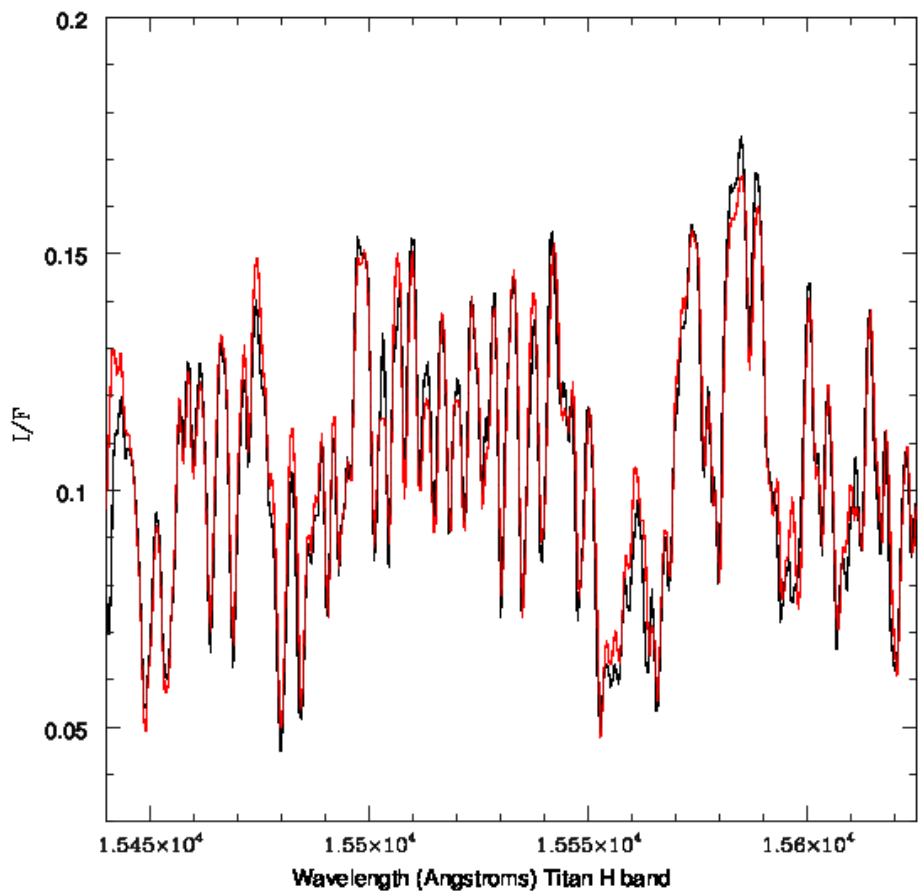}
\caption{GNIRS spectrum of Titan (black) coinciding with the region of lower resolution NIFS spectrum from Figure~\ref{figure1}. VSTAR model is overlaid in red.}
\label{figure3}
\end{figure}

In June 2011 we observed Titan with the Gemini Near Infra-Red Spectrometer (GNIRS) on the Gemini
North 8m telescope, which allowed spectroscopy with the increased
resolving power of R$\sim$18000. The data were reduced by using the Gemini IRAF GNIRS package. We
removed the solar and telluric features resulting form the Earth's atmosphere by dividing the extracted Titan
spectra by the spectrum of a G-type standard star, which was observed immediately after or
before the target object.  We used the same model of Titan's atmosphere adopted for the low resolution
NIFS spectrum, with adjustments in albedo and atmospheric opacity parameters, which are
subject to seasonal changes. The D/H ratio was fixed at the level derived in \cite{Bailey2012}.
The observed GNIRS spectrum is shown in Figures~\ref{figure2} and~\ref{figure3} in black with
the VSTAR model overplotted in red. The region corresponding to that shown in Figure~\ref{figure1}, which
contains the absorption due to the 3$\nu_{2}$ band of CH$_{3}$D, is visible in
Figure~\ref{figure3}. Although there is a close agreement between our data and model, the match
of the features present in both spectra can be improved, as suggested by new modelling of
this region undertaken by Campargue et al. \cite{Campargue2012}. We are currently exploring the
sensitivity of this result to refinements in additional parameters, such as changes in methane mixing ratio in Titan's lower atmosphere.

\section*{Discussion and conclusions} 

In this work, we detail the progress of our investigations into the D/H ratio contained
within objects in the outer regions of the Solar system. Here, we have concentrated
on the deuteration of methane in the atmosphere of Saturn's largest satellite, Titan. Two
different models exist that attempt to explain the origin of methane in the primordial
Titan. Both models feature the progressive enrichment of deuterium in methane in the Solar
nebula - and therefore predict that the level of deuteration in Titan's atmosphere will
depend on the heliocentric distance at which the planetesimals incorporated into Titan
formed, along with any subsequent processing that they experienced in the circum-Saturnian
disk and further processing over the course of Titan's life. 

Both models for the origin of Titan's methane argue that the ongoing enrichment of deuterium in methane would be far slower once
the local temperature fell below 200K, suggesting that the bulk of processing experienced by the
methane likely happened in the proto-planetary disk, prior to Titan's formation. As such, the
Titanian D/H ratio is almost certainly strongly diagnostic of its true formation location, and
therefore a strong indication of the formation location of Saturn itself. That, in turn, will
provide an important datum for studies that are attempting to dynamically reconstruct the
migration of the giant planets by considering their influence on the various small body
reservoirs observed in the outer Solar system (e.g. \cite{Lyk09, JHASSC}). If Saturn formed at
its current location of 9.5 AU from the Sun, and Titan accreted out of the icy planetesimals
around this area, the expected deuterium enrichment factor in Titan $f$ would be of the order of
5 times the initial proto-solar value (see figure~\ref{DHWater}). If the value of  
$26\times10^{-6}$ obtained for Jupiter \cite{Mahaffy1998} is representative of the protosolar value we
predict $\sim130 \times10^{-6}$ for an object formed at this location. The value
($143\pm16)\times10^{-6}$ obtained from the NIFS data \cite{Bailey2012} is consistent 
with this scenario. However, more accurate measurements would help to further constrain the 
formation location and mechanism.

We are currently in the process of analysing data obtained using higher resolution GNIRS for the four giant
planets (Jupiter, Saturn, Uranus and Neptune) in the H and K bands. We intend to apply a
similar modelling technique as presented in this work to derive accurate D/H values for each
of those planets in turn. We will also use our GNIRS observations for the K-band 2$\mu$m
regions to measure the D/H ratio from the CH$_{3}$D absorption due to the $\nu_{5}+3\nu_{6}$
and $2\nu_{3}+2\nu_{6}$ bands, which were detected in Titan's NIFS spectrum (as presented in
\cite{Bailey2011}). However the detailed modelling of this region is dependent on
availability of intensities in these bands, which still await laboratory measurement. 

It seems likely that the D/H ratio in the atmospheres of Jupiter and Saturn will be strongly
influenced by their accretion of hydrogen and deuterated hydrogen directly from the
proto-planetary nebula, and therefore they might be expected to display values similar to
the Solar value (since only a small fraction of their hydrogen will have been accreted in
the form of solid volatile material such as water and methane). By contrast, the
hydrogen/deuterium budget of Uranus and Neptune was most likely primarily accreted in the
form of solids - such as water, methane and ammonia. The deuteration of these planets, then,
will likely strongly reflect the accretion location of their volatile budgets - and
therefore, the location at which those planets accreted the bulk of their mass. The
situation is perhaps even more clear cut for the regular satellites of the giant planets
(such as Titan), whose volatile budget will have been accreted entirely in the form of solid
planetesimals. However, it is possible that the volatiles accreted to these bodies underwent
further processing in circum-planetary disks around their host planet prior to accreting to
form the satellites we observe today. 

Our future studies of deuteration in the outer Solar system will help to answer a variety of
outstanding questions on the formation and evolution of our Solar system, and, taken in
concert with independent dynamical tests of planetary migration, should allow us to finally
determine the true architecture of our Solar system {\it prior to the migration of its giant
planets.}

%
%




\section*{Acknowledgments}

Based on observations obtained at the Gemini Observatory, which is operated by the Association of Universities for
    Research in Astronomy, Inc., under a cooperative agreement with the
    NSF on behalf of the Gemini partnership: the National Science
    Foundation (United States), the Science and Technology Facilities
    Council (United Kingdom), the National Research Council (Canada),
    CONICYT (Chile), the Australian Research Council (Australia),
    Minist\'{e}rio da Ci\^{e}ncia, Tecnologia e Inova\c{c}\~{a}o (Brazil)
    and Ministerio de Ciencia, Tecnolog\'{i}a e Innovaci\'{o}n Productiva
    (Argentina). The observations were obtained under the program GN-2011A-Q-36.



%

\end{document}